\documentclass{PoS}

\newcommand{\Pom}{\mathbb{P}}
\newcommand{\Ode}{\mathbb{O}}

\title{Searching for odderon in exclusive reactions:\\
   $p p \to p p p {\bar p}$, $p p \to p p \phi \phi$ and 
   $p p \to p p \phi$}

\ShortTitle{Searching for odderon in exclusive reactions}

\author{\speaker{Antoni Szczurek}\thanks{A footnote may follow.}\\
        Institute of Nuclear Physics, PAN, Krak\'ow\\
        E-mail: \email{antoni.szczurek@ifj.edu.pl}}

\author{Piotr Lebiedowicz\\
        Institute of Nuclear Physics, PAN, Krak\'ow\\
        E-mail: \email{piotr.lebiedowicz@ifj.edu.pl}}

\abstract{There seem to be recently an evidence for $C =$ -1 exchanges in
$p p$ and $p \bar p$ elastic scattering at high energies. 
The analysis there is difficult as the two processes were not measured 
at the same (large) energies.
Here we discuss three different exclusive processes given in 
the title as a possible source of information for odderon exchange.
A sketch of the formalism is presented for each of the reactions.
We consider low energy processes measured in the past by the WA102
collaboration and try to make predictions for the LHC.
We discuss possible evidences at the low energies and
try to make suggestions for the LHC.} 

\FullConference{XXVII International Workshop on Deep-Inelastic Scattering and Related Subjects - DIS2019\\
		8-12 April, 2019\\
		Torino, Italy}

\begin{document}

\section{Introduction}

The odderon exchange became recently topical again.
So far there is no unambiguous evidence for the odderon -- 
the $C = -1$ partner of the pomeron \cite{LN1973}.
For a theoretical review of the odderon see e.g.~\cite{Ewerz_review}.
Recent analysis of the elastic scattering by the TOTEM collaboration 
\cite{TOTEM_old,Antchev:2018rec} 
allow for the (soft) odderon exchange interpretation of the data 
\cite{MN2018}.

In this talk we discussed three processes mentioned in the title
\cite{LNS2019,LNS2018,LNS2019_phi}.
We try te estimate upper limit of the size of odderon exchanges in these
exclusive processes.

\section{A sketch of formalism}

We discuss shortly the three considered here reactions.

\subsection{$p p \to p p \phi \phi$}

In Fig.\ref{fig:diagrams_phiphi} we show conventional processes
leading to the exclusive production of two $\phi$ mesons. 

\begin{figure}
\begin{center}
(a)\includegraphics[width=5.5cm]{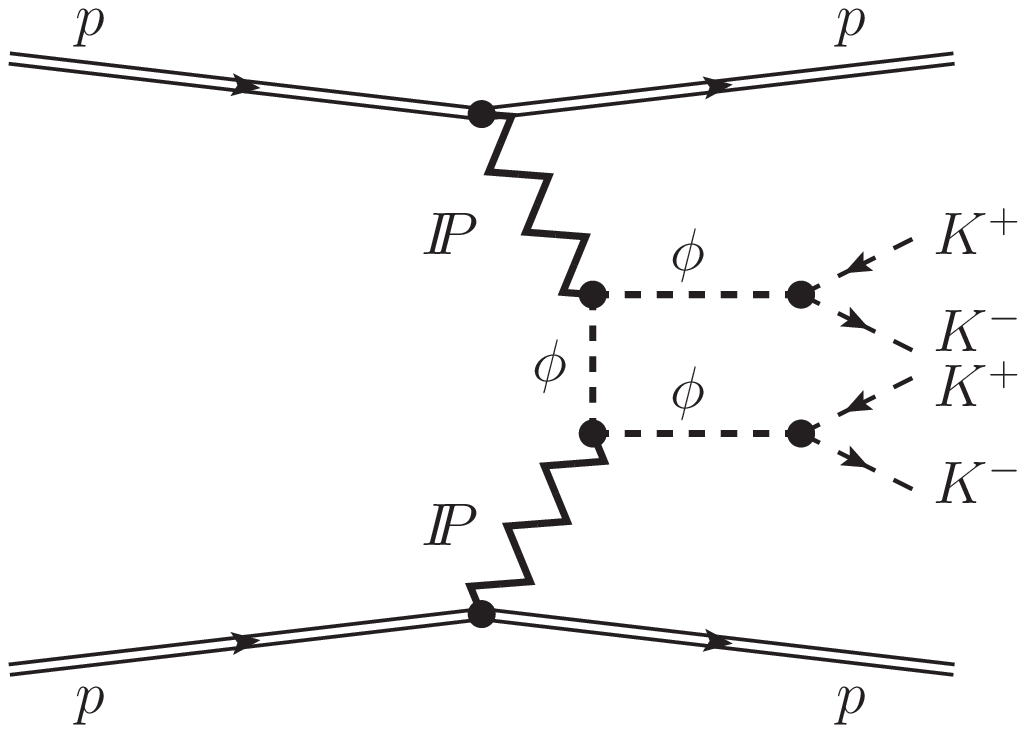}    
(b)\includegraphics[width=5.5cm]{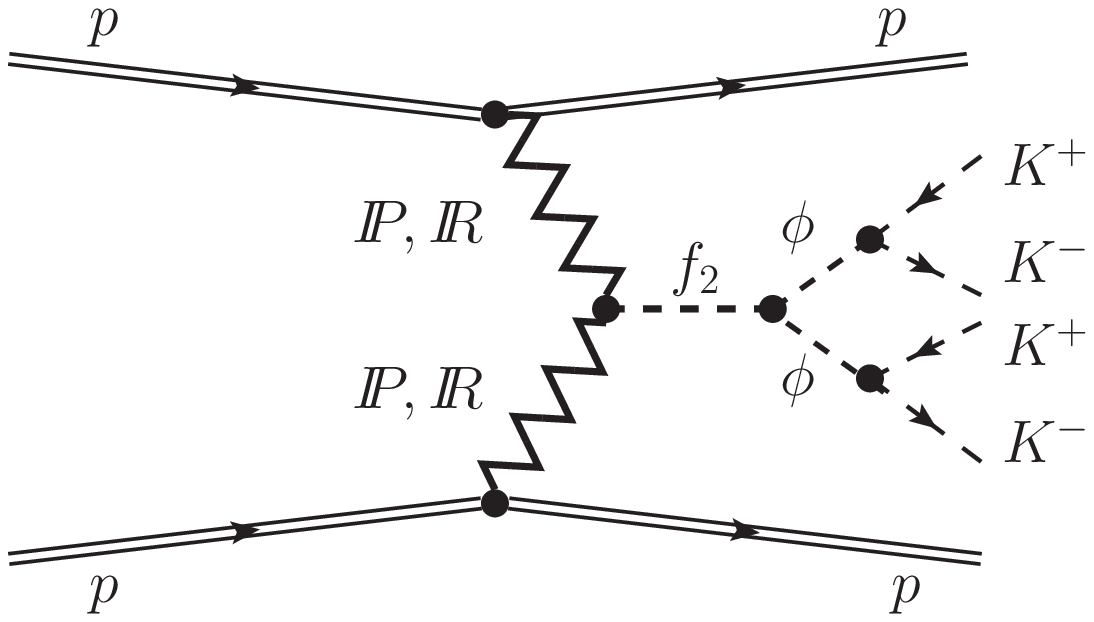}  
  \caption{\label{fig:4K_diagrams}
The ``Born level'' diagrams for double-pomeron/reggeon
central exclusive $\phi \phi$ production 
and their subsequent decays into $K^+ K^- K^+ K^-$ in proton-proton collisions.
In (a) we have the continuum $\phi\phi$ production,
in (b) $\phi\phi$ production via an $f_{2}$ resonance.
Other resonances, e.g. of $f_{0}$- and $\eta$-type, can also contribute here.
}
\end{center}
\label{diagrams_phiphi}
\end{figure}

Let us sketch the formalism to descibe the conventional
processes.

For the $f_{2} \phi \phi$ vertex we take the following ansatz,
in analogy to the $f_{2}\gamma\gamma$ vertex:
\begin{eqnarray}
i\Gamma^{(f_{2} \phi \phi)}_{\mu \nu \kappa \lambda}(p_{3},p_{4}) &=&
i\frac{2}{M_{0}^{3}} g'_{f_{2} \phi \phi}\,  
\Gamma^{(0)}_{\mu \nu \kappa \lambda}(p_{3},p_{4})\,
F'^{(f_{2} \phi \phi)}(p_{34}^{2}) \nonumber \\
&&- i \frac{1}{M_{0}} g''_{f_{2} \phi \phi}\,
\Gamma^{(2)}_{\mu \nu \kappa \lambda}(p_{3},p_{4})\,
F''^{(f_{2} \phi \phi)}(p_{34}^{2})\,,
\label{vertex_f2phiphi}
\end{eqnarray}  
with $M_{0} = 1$~GeV and dimensionless coupling constants 
$g'_{f_{2} \phi \phi}$ and $g''_{f_{2} \phi \phi}$ 
being free parameters.
Two free couplings are allowed in general.

The propagator for $\phi$ exchange in diagram (a) must be modified
(compared to the corresponding one for meson exchange) at higher
$\phi \phi$ subsytem energies. In \cite{LNS2019} we proposed the following
parametrization:
\begin{eqnarray}
\Delta^{(\phi)}_{\rho_{1}\rho_{2}}(\hat{p}) \to
\Delta^{(\phi)}_{\rho_{1}\rho_{2}}(\hat{p}) 
\left( \exp (i \phi(s_{34}))\, \frac{s_{34}}{s_{0}} \right)^{\alpha_{\phi}(\hat{p}^{2})-1} \,,
\label{reggeization}
\end{eqnarray}
where we take $s_{0} = 4 m_{\phi}^{2}$ and 
$\alpha_{\phi}(\hat{p}^{2}) = \alpha_{\phi}(0) + \alpha'_{\phi} \,\hat{p}^{2}$ 
with $\alpha_{\phi}(0) = 0.1$ from \cite{Collins_book} 
and $\alpha'_{\phi}=0.9$~GeV$^{-2}$.

The different considered processes are added coherently and therefore 
interfere.
In order to have a correct phase behaviour 
we introduced the function $\exp (i \phi(s_{34}))$ with
\begin{eqnarray}
\phi(s_{34}) =\frac{\pi}{2}\exp\left(\frac{s_{0}-s_{34}}{s_{0}}\right)-\frac{\pi}{2}\,
\label{reggeization_aux}
\end{eqnarray}
which role is to interpolate between meson physics close to 
the $\phi \phi$ threshold, $s_{34} = 4 m_{\phi}^2$, and 
Regge physics at high energies.
Another prescription was considered in \cite{LNS2019} in addition.

\begin{figure}
\begin{center}
(a)\includegraphics[width=5.5cm]{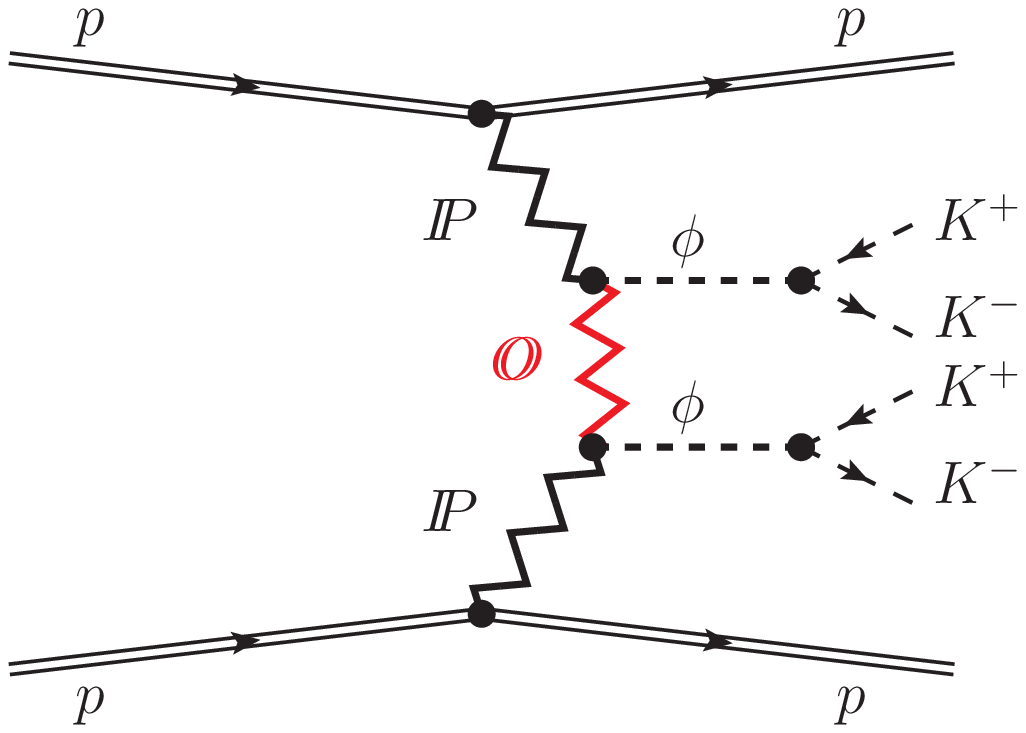}    
(b)\includegraphics[width=5.5cm]{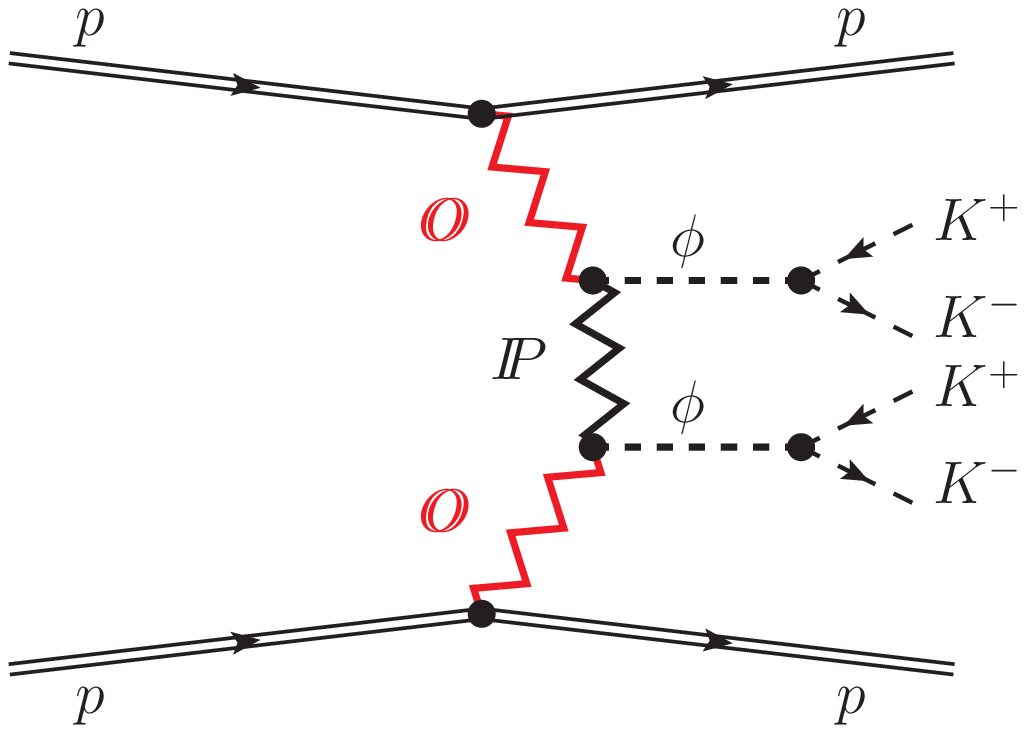}  
  \caption{\label{fig:diagrams_odderon}
The Born level diagrams for diffractive production of 
a $\phi$-meson pair with one and two odderon exchanges.}
\end{center}
\end{figure}

In addition to the processes discussed above we considered processes
shown in Fig.\ref{fig:diagrams_odderon}. We think that the contribution
of the diagram with two odderon exchanges is much smaller than that for
the diagram with one odderon exchange. Therefore we include only
contribution of the diagram with odderon exchange in the middle
of the left diagram.

Our ansatz for the effective propagator of $C = -1$ odderon follows 
\cite{EMN2014}
\begin{eqnarray}
&&i \Delta^{(\Ode)}_{\mu \nu}(s,t) = 
-i g_{\mu \nu} \frac{\eta_{\Ode}}{M_{0}^{2}} 
( s \alpha'_{\Ode})^{\alpha_{\Ode}(t)-1}\,,
\label{A12} \\
&&\alpha_{\Ode}(t) = \alpha_{\Ode}(0)+\alpha'_{\Ode}\,t\,,
\label{A13}
\end{eqnarray}
where we have $M_{0}^{-2} = 1$~(GeV)$^{-2}$ for dimensional reasons.
Furthermore, we shall assume representative values for the odderon 
parameters 
\begin{eqnarray}
&&\eta_{\Ode} = -1\,, \qquad
\alpha_{\Ode}(0) = 1.05\,,\qquad
\alpha'_{\Ode} = 0.25 \; \mathrm{GeV}^{-2}\,.
\label{A14}
\end{eqnarray}

For the $\Pom \Ode \phi$ vertex we use an ansatz
analogous to the $\Pom \rho \rho$ vertex. 
We get then, orienting the momenta of the $\Ode$ and the $\phi$
outwards, the following formula:
\begin{eqnarray}
i\Gamma^{(\Pom \Ode \phi)}_{\mu \nu \kappa \lambda}(k',k) =
i F^{(\Pom \Ode \phi)}( (k+k')^{2},k'^{2},k^{2})
\left[ 2\,a_{\Pom \Ode \phi}\, 
\Gamma^{(0)}_{\mu \nu \kappa \lambda}(k',k)
- b_{\Pom \Ode \phi}\,
\Gamma^{(2)}_{\mu \nu \kappa \lambda}(k',k) \right].\qquad
\label{A15}
\end{eqnarray}  
%

\subsection{$p p \to p p \phi$}

For single $\phi$ production we include processes shown in 
Fig.3.

\begin{figure}
\begin{center}
\includegraphics[height=3.5cm]{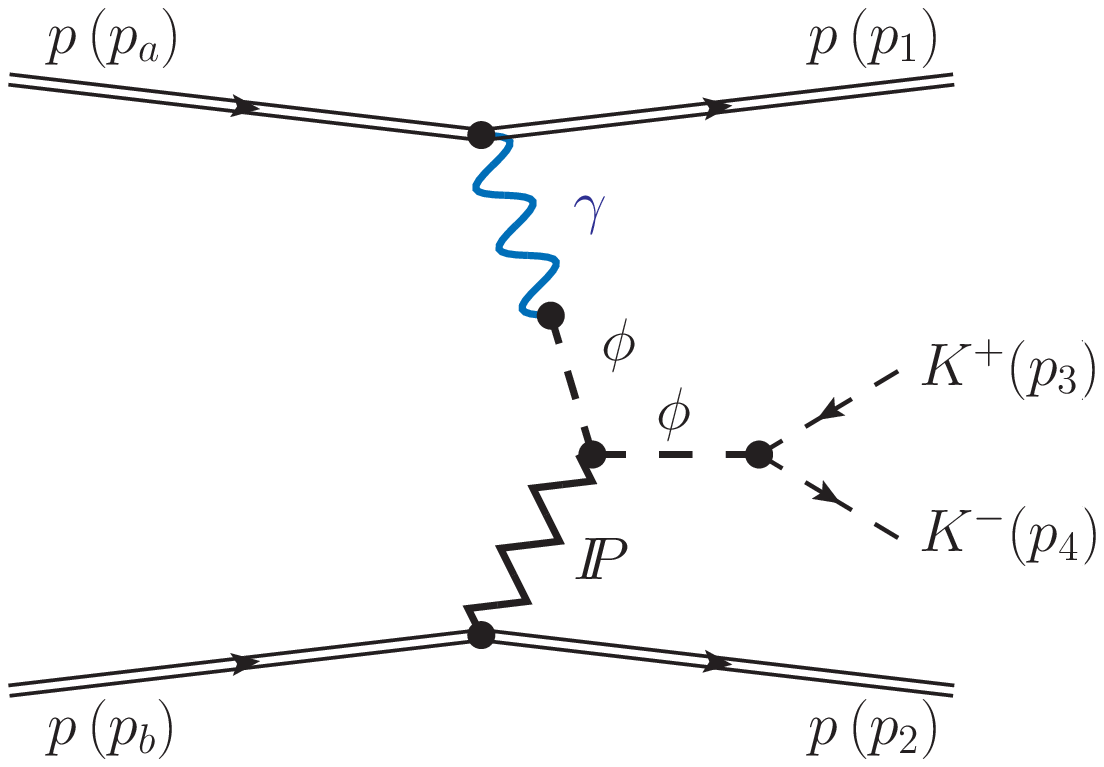}
\includegraphics[height=3.5cm]{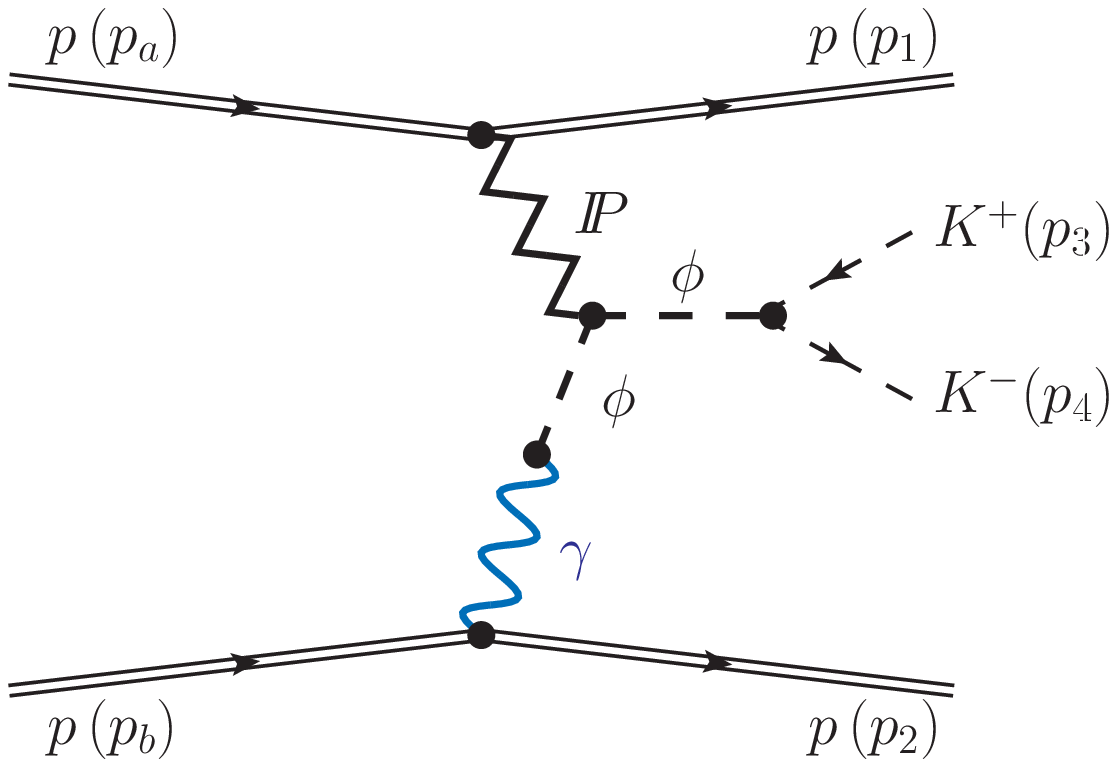}\\
\includegraphics[height=3.5cm]{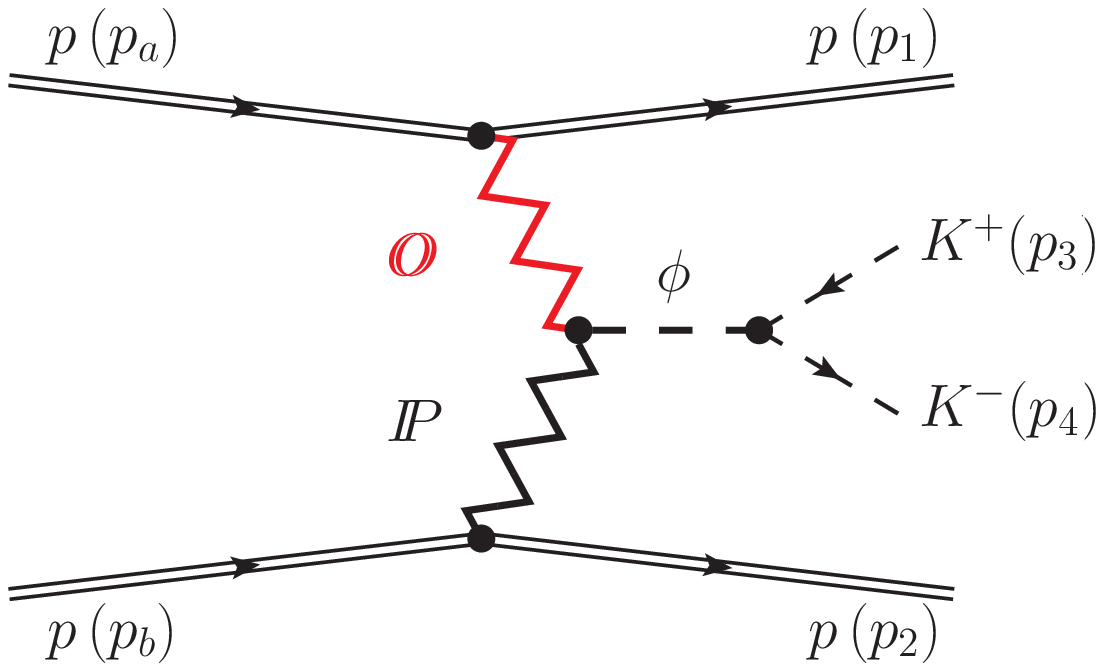}
\includegraphics[height=3.5cm]{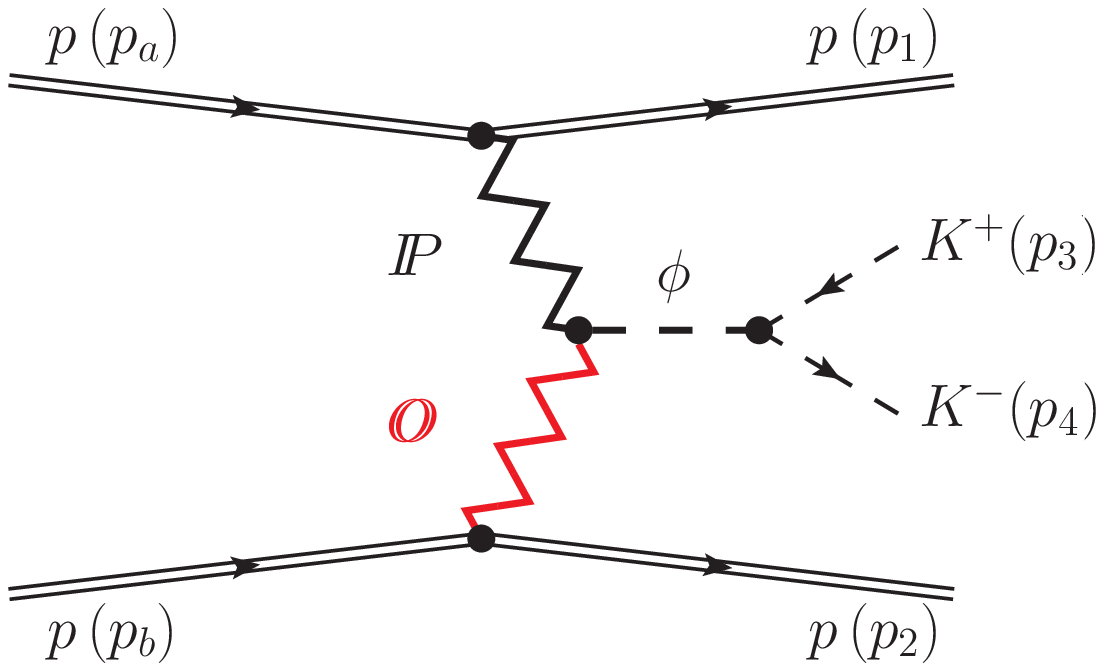}
\caption{Some diagrams included in the analysis of single $\phi$ production.}
\end{center}
\label{fig:diagrams_singlephi}
\end{figure}

The Odderon exchange contributions (lower row) may modify 
the photon-exchange contribution (upper low). 
More processes will be considered in \cite{LNS2019}.

\subsection{$p p \to p p p \bar p$}

Several mechanisms of central $p \bar p$ production were discussed
in \cite{LNS2018} and the formalism how to calculate relevant 
diffractive processes (continuum and resonances) was given there.

The exchange of $C =$ -1 objects leads to specific asymmetries 
discussed in \cite{LNS2018}.
In two dimensions (e.g. $\eta_1, \eta_2$) we can define the asymmetry:

\begin{eqnarray}
\widetilde{A}^{(2)}(\eta,\eta')=\frac{\frac{d^{2}\sigma}{d\eta_{3} d\eta_{4}}(\eta,\eta')
                                 -\frac{d^{2}\sigma}{d\eta_{3} d\eta_{4}}(\eta',\eta)}
                                 {\frac{d^{2}\sigma}{d\eta_{3} d\eta_{4}}(\eta,\eta')
                                 +\frac{d^{2}\sigma}{d\eta_{3} d\eta_{4}}(\eta',\eta)}\,.
\label{asymmetry_3}
\end{eqnarray}

As discussed in \cite{LNS2018} the $C =$-1 reggeon exchanges lead 
to such asymmetries. The odderon exchange also leads to asymmetries.
How big are such asymmetries was discussed in detail in \cite{LNS2018}.

\section{Results}

\begin{figure}[!ht]
\includegraphics[width=0.42\textwidth]{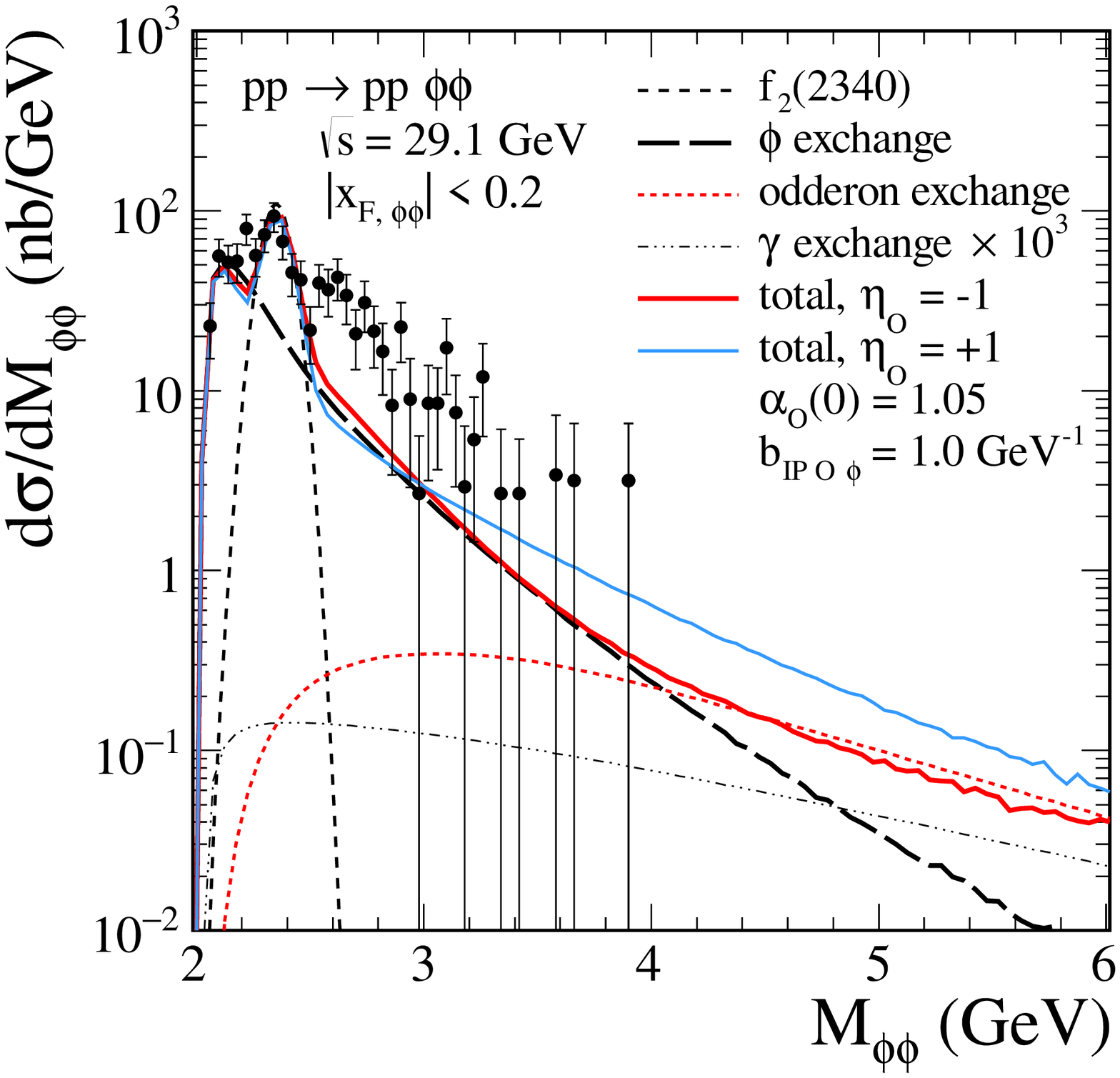}
\includegraphics[width=0.42\textwidth]{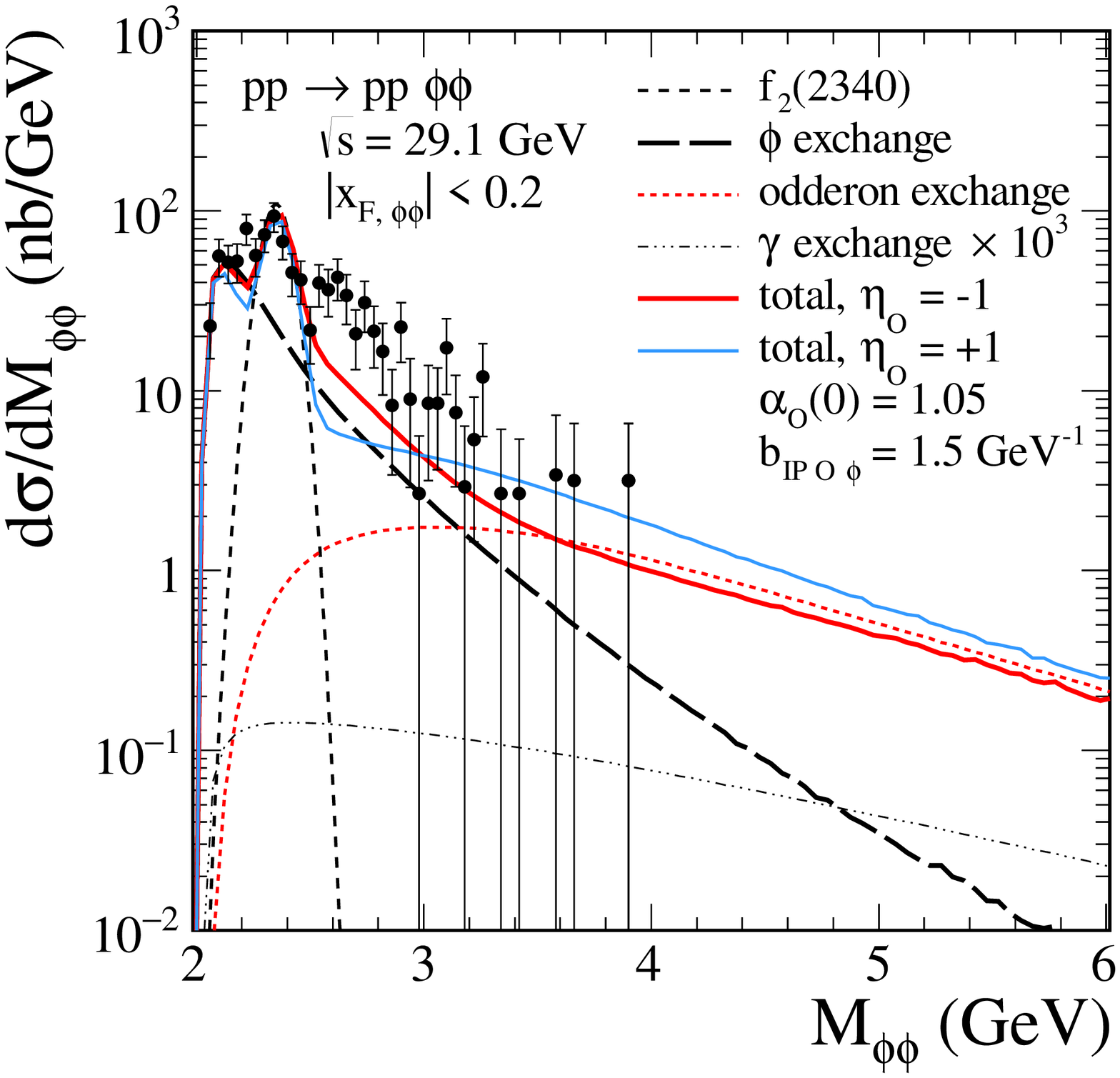}
\caption{\label{fig:odderon_WA102}
Invariant mass distributions for the central production of $\phi \phi$ at $\sqrt{s} = 29.1$~GeV
and $|x_{F,\phi \phi}| \leqslant 0.2$ together with the WA102 data 
\cite{WA102_phiphi} are shown.
The black long-dashed line corresponds to the $\phi$-exchange contribution 
and the black dashed line corresponds to the $f_{2}(2340)$ contribution.
The black dot-dashed line corresponds to the $\gamma$-exchange contribution 
enlarged by a factor $10^{3}$.
The red dotted line represents the odderon-exchange contribution
for $a_{\Pom \Ode \phi}=0$, $b_{\Pom \Ode \phi}= 1.0$~GeV$^{-1}$ (the left panel) 
and for $a_{\Pom \Ode \phi}=0$, $b_{\Pom \Ode \phi}= 1.5$~GeV$^{-1}$ (the right panel).
The coherent sum of all terms is shown by the red and blue solid lines
for $\eta_{\Ode} = -1$ and $\eta_{\Ode} = +1$, respectively.
Here we take $\alpha_{\Ode}(0) = 1.05$.
The absorption effects are included in the calculations.}
\end{figure}

It is very difficult to describe the WA102 data for $p p \to p p \phi \phi$
reaction including resonances and $\Pom \Pom$ mechanism only
\cite{LNS2019}.
Inclusion of the odderon exchange improves the description of the data.
The result of our analysis is shown in Fig.\ref{fig:odderon_WA102}.

\begin{figure}[!ht]
\begin{center}
\includegraphics[width=0.38\textwidth]{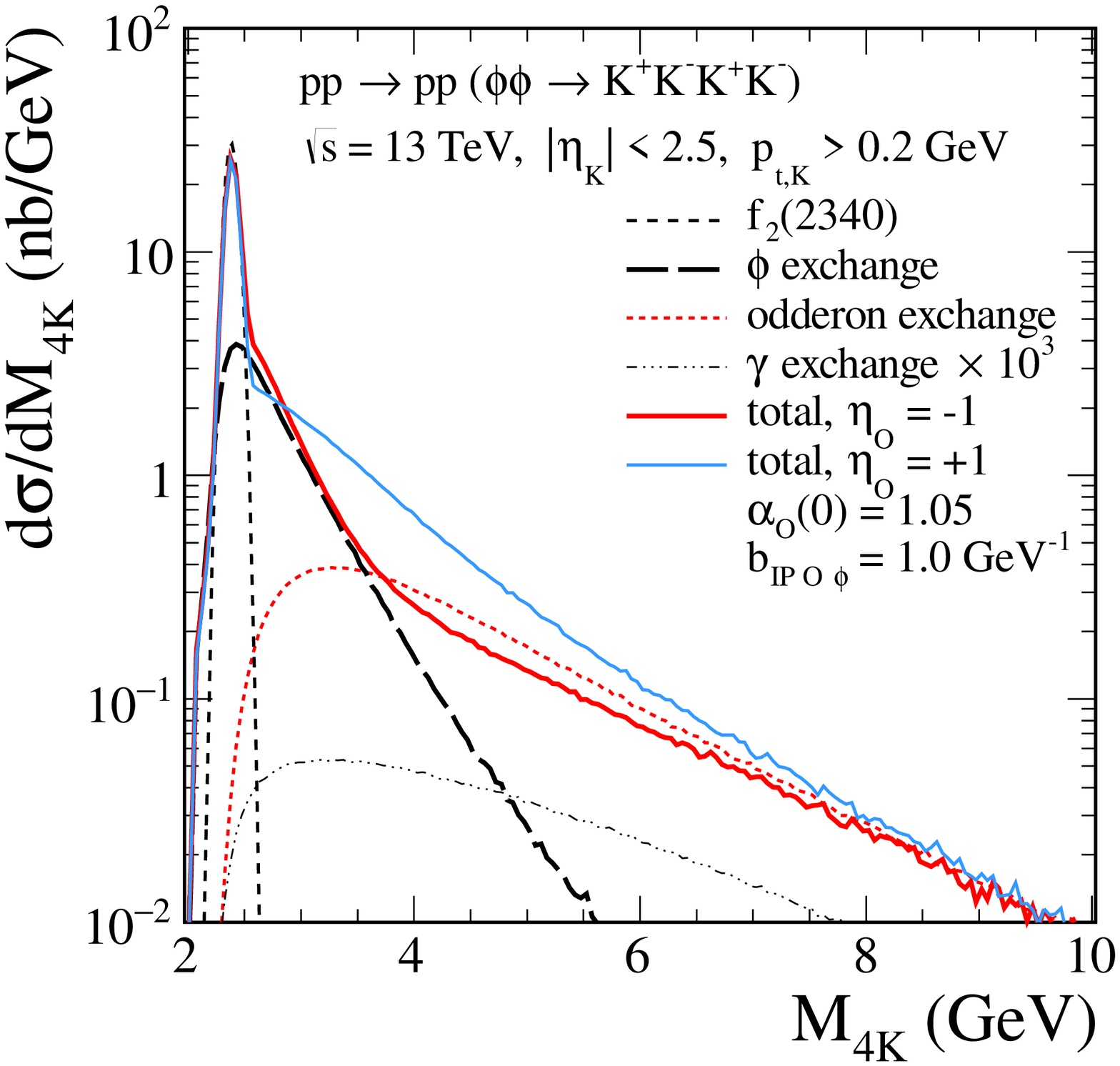}
\includegraphics[width=0.38\textwidth]{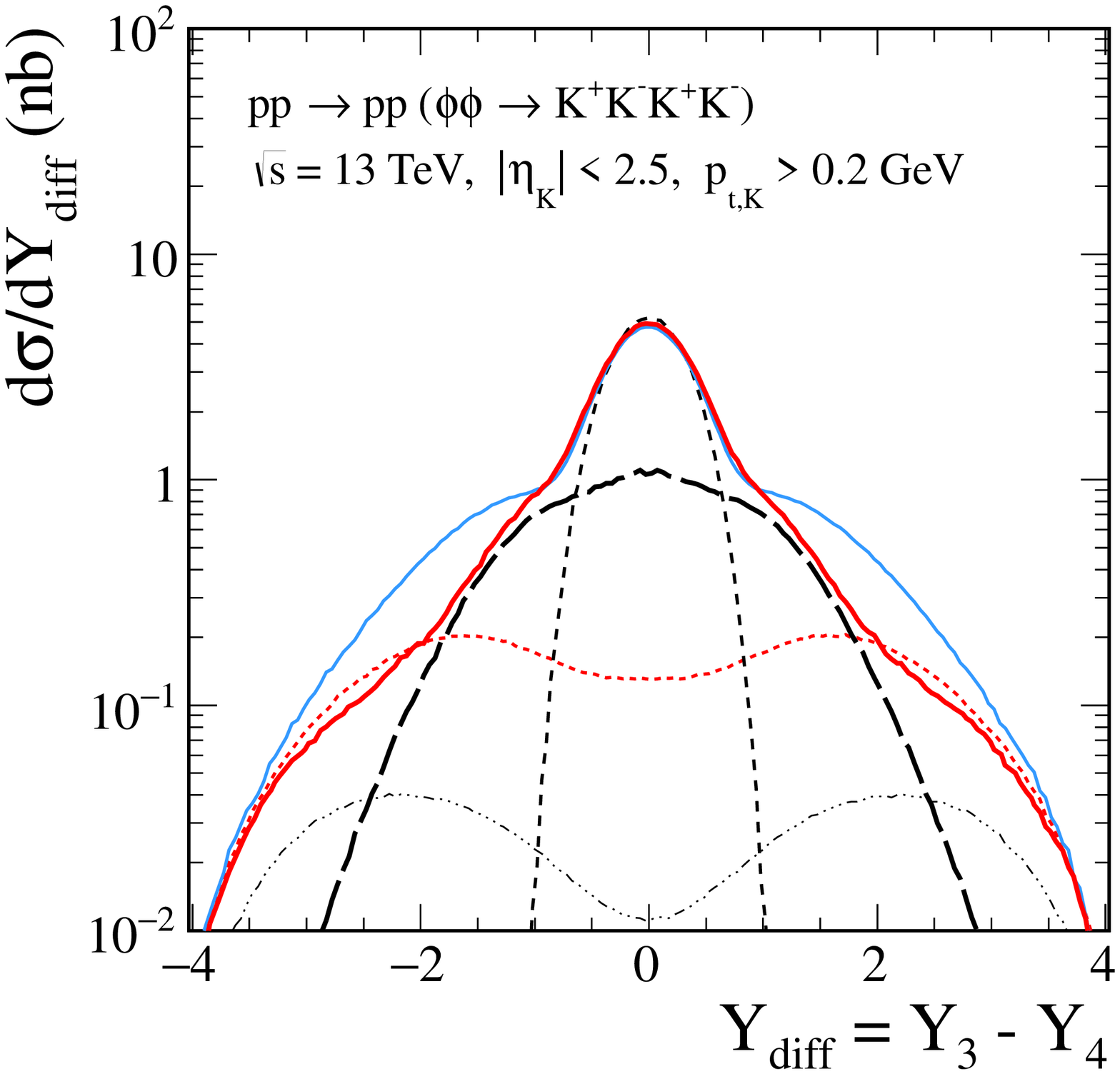}\\
\includegraphics[width=0.38\textwidth]{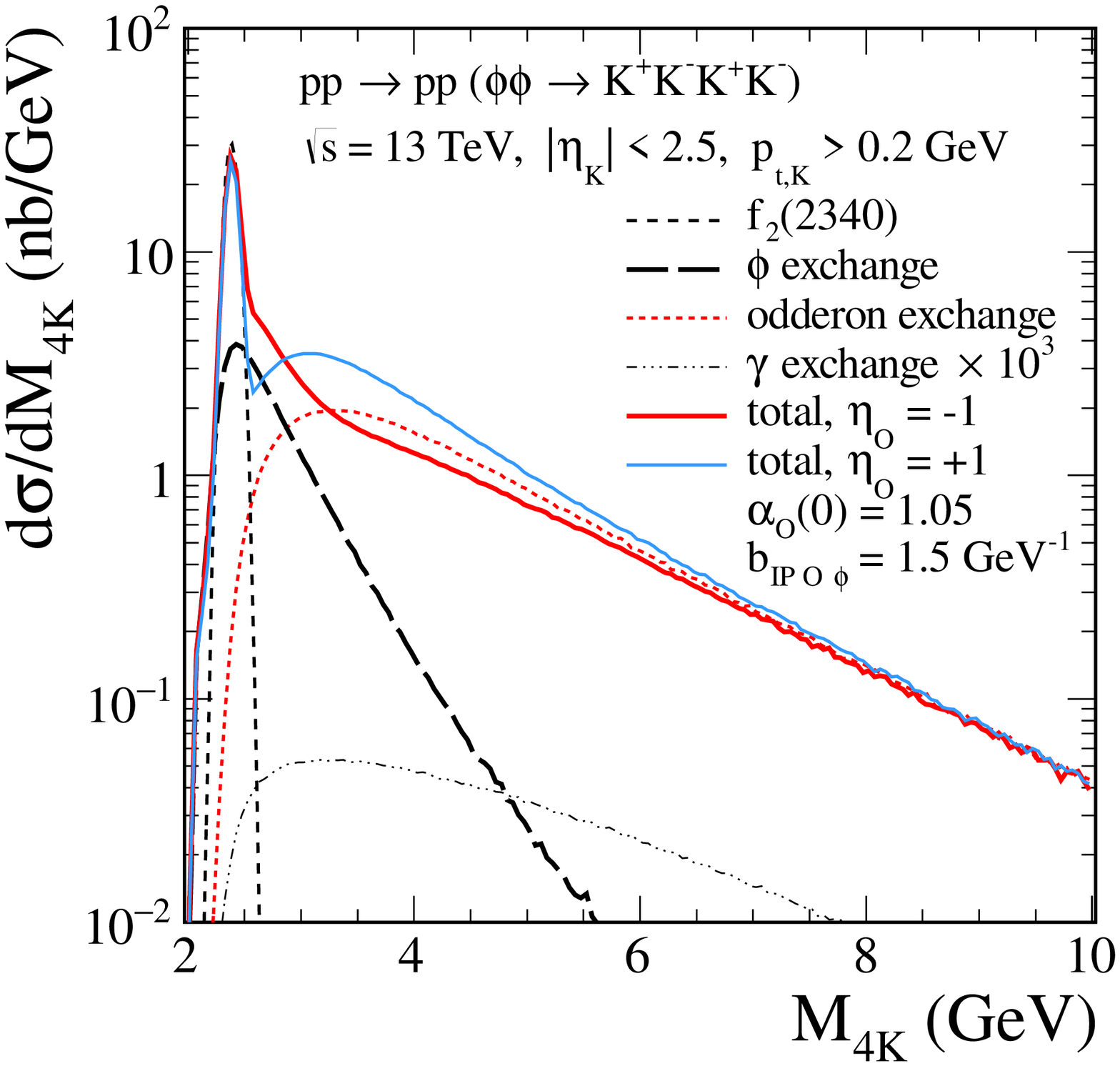}
\includegraphics[width=0.38\textwidth]{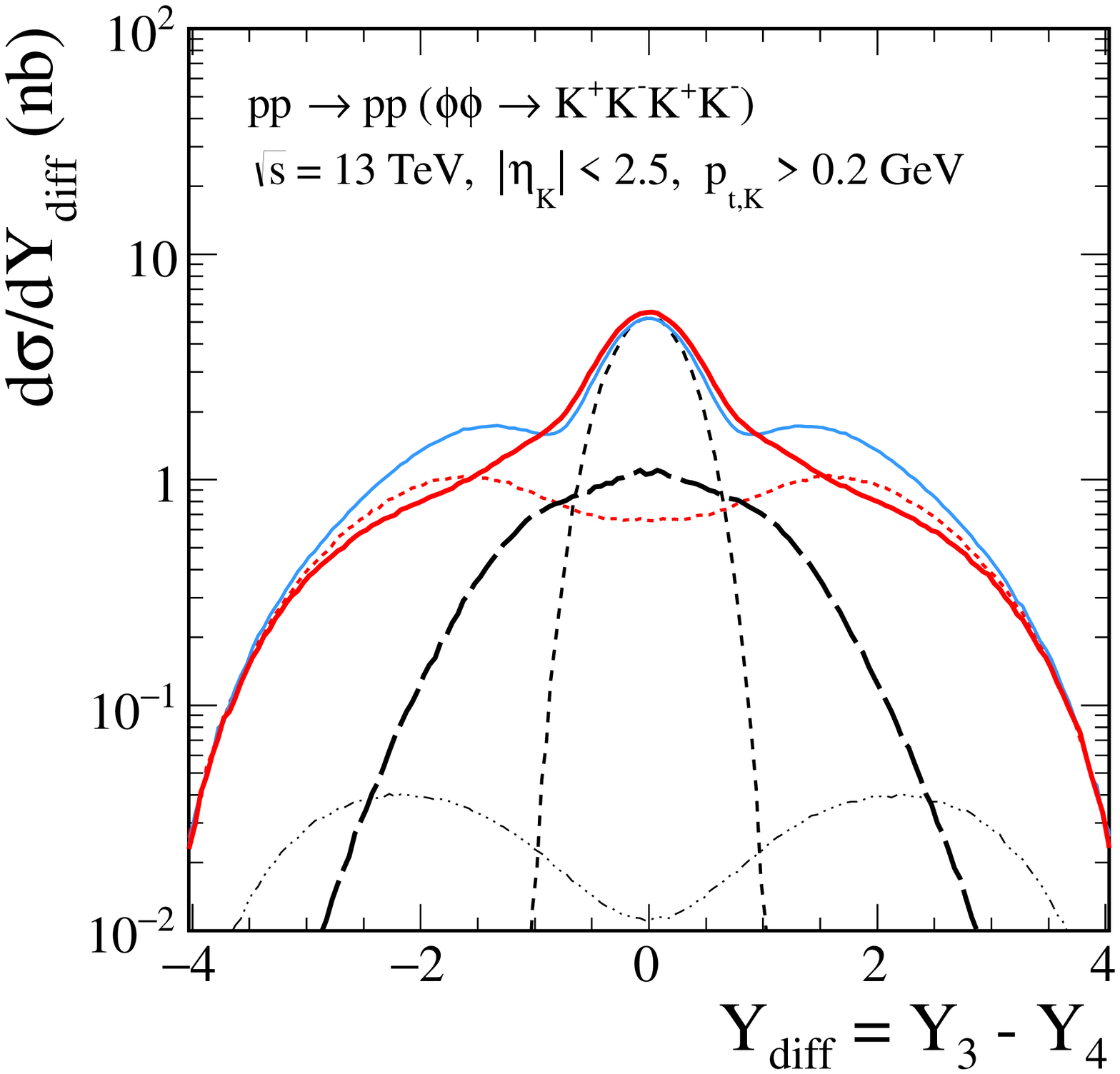}
\caption{\label{fig:odderon_LHC}
The distributions in ${\rm M}_{4K}$ (the left panels)
and in $\rm{Y_{diff}}$ (the right panels)
for the $pp \to pp (\phi \phi \to K^{+}K^{-}K^{+}K^{-})$ reaction 
calculated for $\sqrt{s} = 13$~TeV and $|\eta_{K}| < 2.5$, 
$p_{t, K} > 0.2$~GeV.
The red and blue solid lines correspond to the complete results with
$\eta_{\Ode} = -1$ and 
$\eta_{\Ode} = +1$, respectively.
The results for $b_{\Pom \Ode \phi}= 1.0$~GeV$^{-1}$ (the top panels)
and for $b_{\Pom \Ode \phi}= 1.5$~GeV$^{-1}$ (the bottom panels)  
are presented.
The absorption effects are included in the calculations.}
\end{center}
\end{figure}

Having fixed the parameters of our quasi fit to the WA102 data
we wish to show our predictions for the LHC.
In Fig.~\ref{fig:odderon_LHC} we show the results
for the ATLAS experimental conditions ($|\eta_{K}| < 2.5$, $p_{t, K} >
0.2$~GeV).
The distribution in four-kaon invariant mass is shown in the left panels
and the difference in rapidity between the two $\phi$ mesons in the 
right panels.

\begin{figure}[!ht]
\begin{center}
\includegraphics[width=0.42\textwidth]{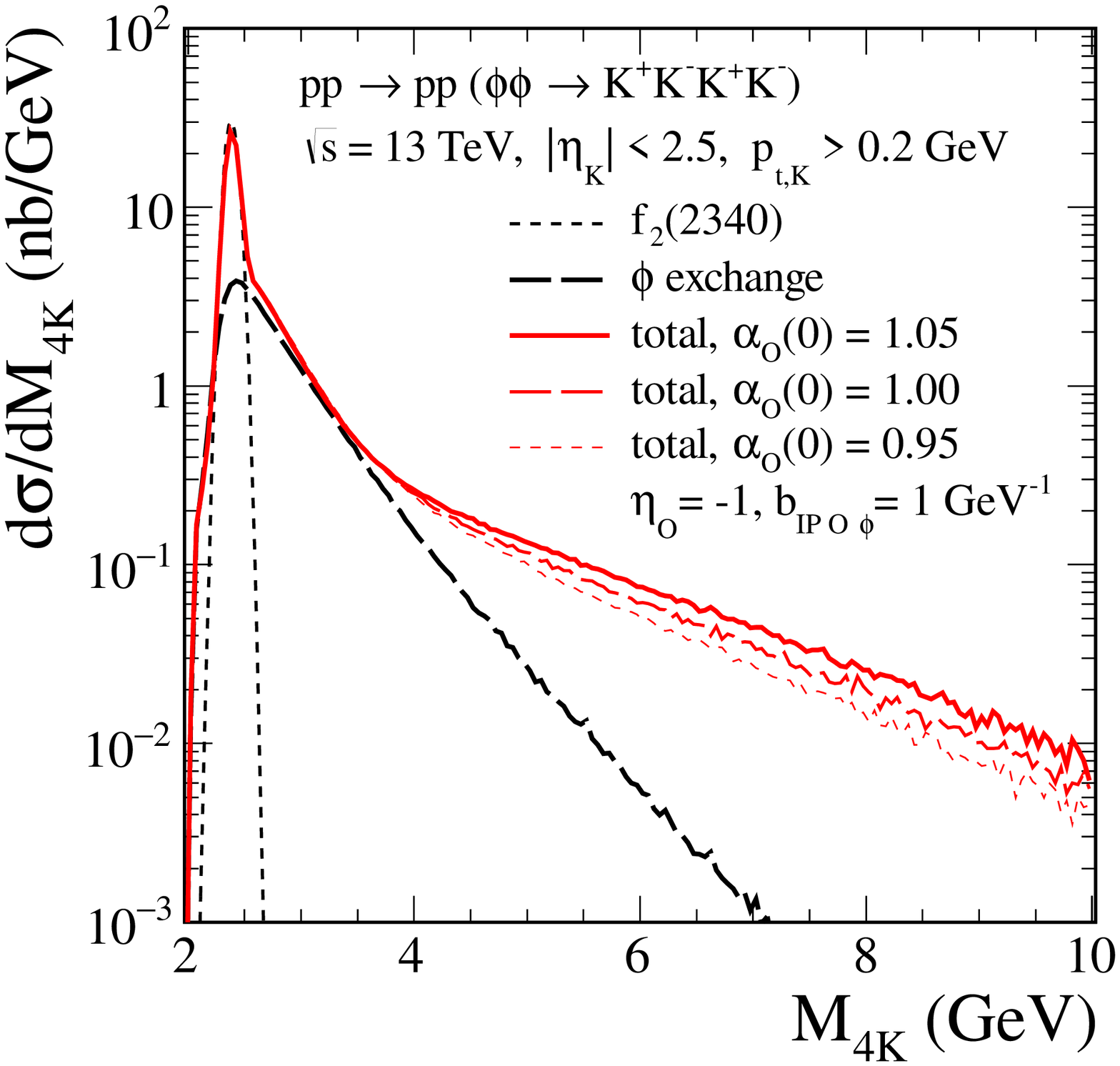}
\includegraphics[width=0.42\textwidth]{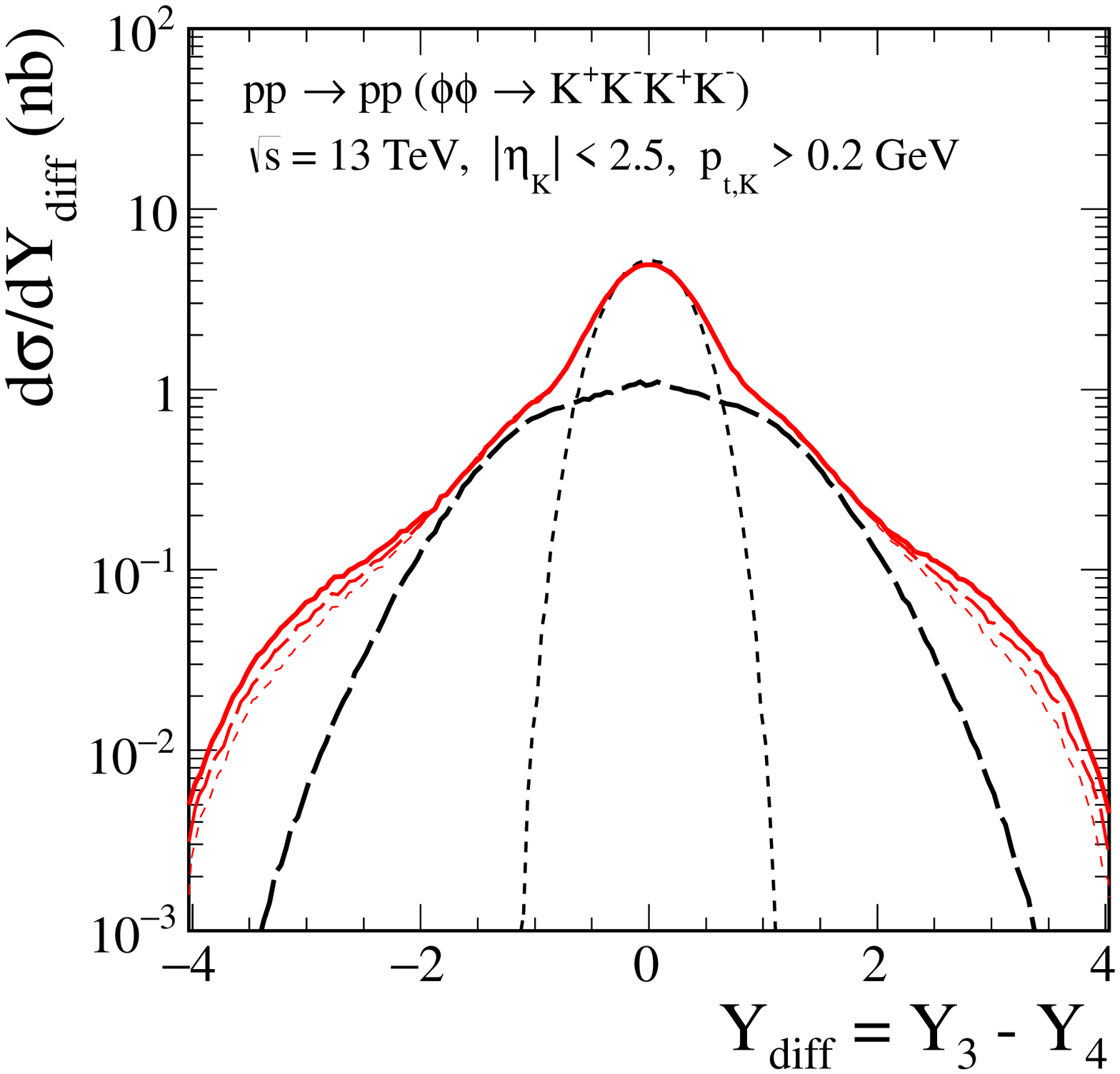}
\caption{\label{odderon_LHC_aux2}
\small
The complete results for $\sqrt{s} = 13$~TeV and $|\eta_{K}| < 2.5$, 
$p_{t, K} > 0.2$~GeV are shown.
Here we show results for $\eta_{\Ode} = -1$ 
and for various values of the odderon intercept $\alpha_{\Ode}(0)$.
Here we take $a_{\Pom \Ode \phi}= 0$ and $b_{\Pom \Ode \phi}=
1$~GeV$^{-1}$.
Odderon could be visible for $M_{\phi \phi} >$ 6 GeV
and/or for $Y_{diff} >$ 3.
}
\end{center}
\end{figure}

In Fig.~\ref{odderon_LHC_aux2} we show more  
complete result including the odderon exchange with
$\eta_{\Ode} = -1$ and various values of the odderon intercept 
$\alpha_{\Ode}(0)$.

The measurement of large $M_{\phi \phi}$ or $Y_{diff}$ events
at the LHC would therefore suggest presence of the odderon exchange.

We will not present here (in the written version) preliminary results for
the $p p \to p p \phi$ reaction. Some preliminary results were presented
in the talk and details will be presented soon in \cite{LNS2019_phi}.
We wish to mention only here that the inclusion of $\Pom \Ode$ fusion
(see Fig.3)
leads to a sizeable improvement of the description of the rather old
WA102 data \cite{WA102_single_phi} for this reaction.
 
Now we go to the $p p \to p p p {\bar p}$ reaction.
In Fig.7 we show estimated, allowed at present, 
asymmetry defined in detail in \cite{LNS2018}
for two different kinematical conditions specified in the figure description.
Rather small asymmetries are allowed. A measurement of such asymmetries
would require gigantic statistics for the $p p \to p p p {\bar p}$
reaction and may be difficult to reach in planned experiments.

\begin{figure}
\begin{center}
\includegraphics[height=5.5cm]{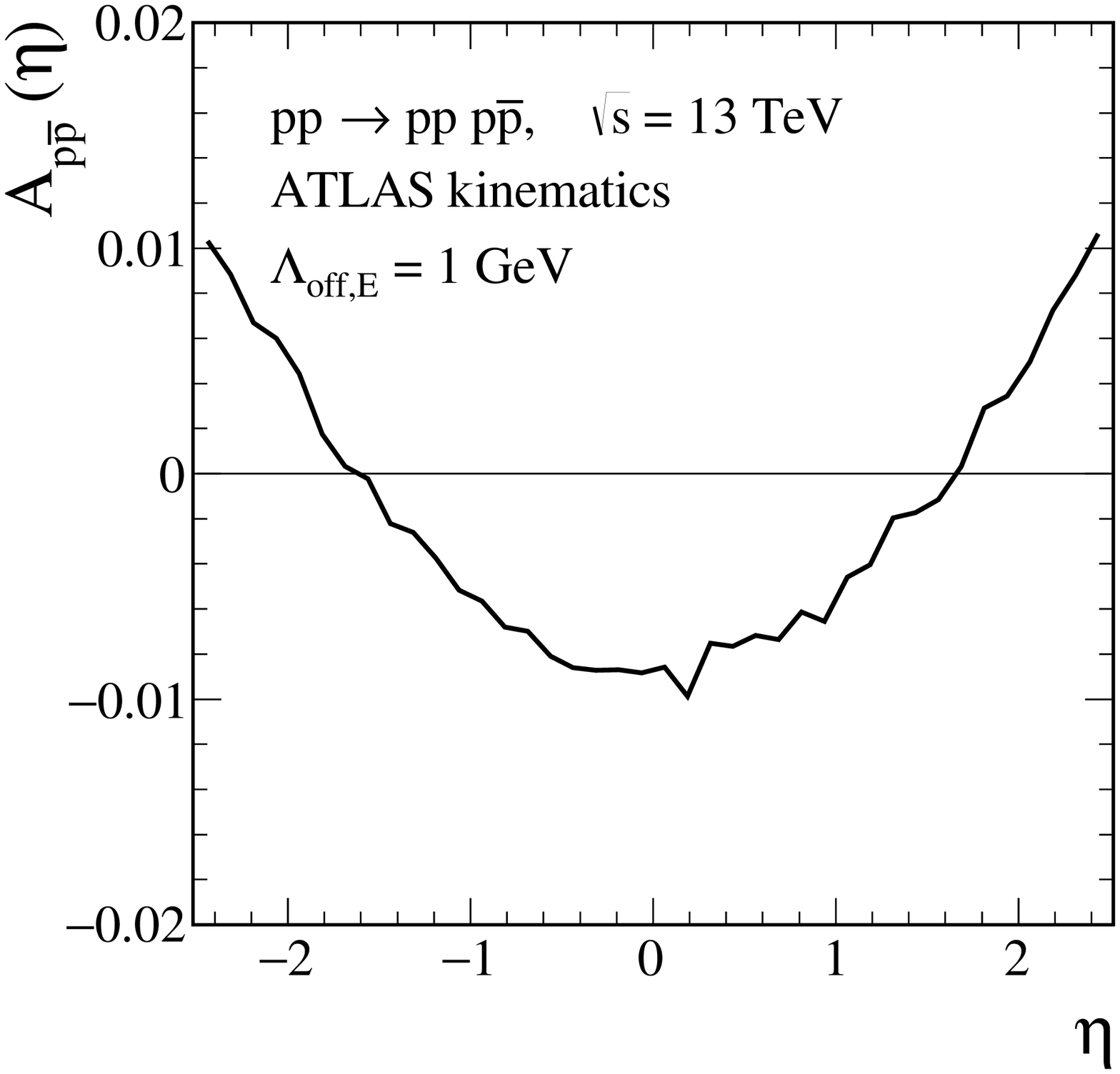}
\includegraphics[height=5.5cm]{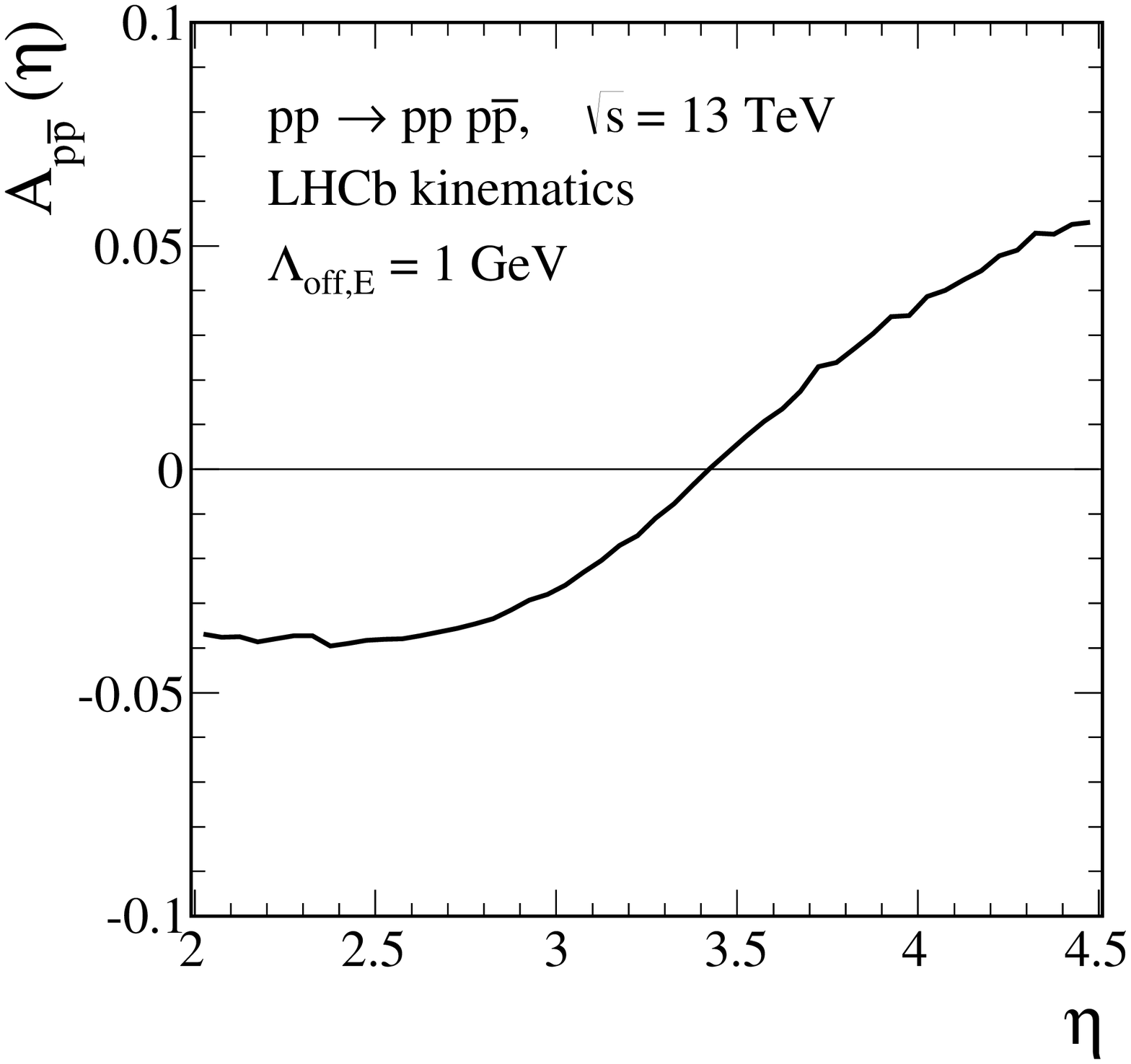}
\caption{\label{fig:asymmetry ppbar}
Allowed asymmetry for $p \bar p$ production due to
  odderon-pomeron exchange for ATLAS and LHCb kinematics.}
\end{center}
\end{figure}

\section{Conclusions}

Our results/presentation can be summarized in the following way:

\begin{itemize}

\item The Regge phenomenology was extended
to $2 \to 3$, $2 \to 4$ and $2 \to 6$ exclusive processes.

\item The tensor pomeron/reggeon model was applied
in all these reactions.

\item At lower energies tensor/vector reggeon exchanges must be included.

\item Three reactions ($p p \to p p \phi \phi$, $p p \to p p \phi$
and $p p \to p p p \bar p$
have been studied in the context of identifying odderon exchange.

\item $p p \to p p \phi \phi$ seems promissing as here the odderon
does not couple to protons. 
An upper limit for the odderon exchange
has been etablished based on the WA102 data.

\item This upper limit for the $\Pom \Ode \to \phi$ coupling
was used for the $p p \to p p \phi$ reaction, together with the TOTEM
estimate. The WA102 data for single $\phi$ production support 
existence of odderon exchange (not shown in the written version).

\item Special asymmetries for centrally produced $p \bar p$ system
have been proposed to identify $C = -1$ exchanges.
The asymmetry caused by subleading reggeon exchanges is probably
considerably larger than that for the odderon exchange which makes 
this reaction difficult in searches for odderon exchanges.

\end{itemize}


\end{document}